\documentclass[pre,aps,reprint,amsmath]{revtex4-1}

\usepackage[T1]{fontenc} 
\usepackage{graphicx}
\usepackage[utf8]{inputenc}
\usepackage{lmodern}
\usepackage{bm}
\usepackage{bbold}
\usepackage{hyperref}

\begin{document}
\title{Active interface polarization is a state function}

\author{Sophie Hermann}
\affiliation{Theoretische Physik II, Physikalisches Institut, Universit{\"a}t Bayreuth, D-95447 Bayreuth, Germany}

\author{Matthias Schmidt}
\affiliation{Theoretische Physik II, Physikalisches Institut, Universit{\"a}t Bayreuth, D-95447 Bayreuth, Germany}
\email{Matthias.Schmidt@uni-bayreuth.de}

\date{25 January 2020, to appear in Physical Review Research (Rap. Comm.)}

\begin{abstract}
We prove three exact sum rules that relate the polarization of active Brownian particles to their one-body current: (i) The total polarization vanishes, provided that there is no net flux through the boundaries, (ii) at any planar wall the polarization is determined by the magnitude of the bulk current, and (iii) the total interface polarization between phase-separated fluid states is rigorously determined by the gas-liquid current difference. This result precludes the influence of the total interface polarization on active bulk coexistence and questions the proposed coupling of interface to bulk.
\end{abstract}

\maketitle
Systems of active Brownian particles (ABPs) consist of thermally diffusing spheres that self propel along an intrinsic direction, which itself undergoes free rotational diffusion \cite{marchetti2013,review2,review3}. ABPs form the prototypical statistical model for active matter \cite{marchetti2013,review2,review3}. In order to characterize the local orientational order, the polarization $\textbf{M}$ is a measure of the strength and direction of the local preferred alignment of the particle orientations. A multitude of relevant situations have been reported in the literature where ABPs display spontaneous polarization effects \cite{speckjack,enculescu, ginot2018,mazza,schmidt2018,utrecht2018, stark2014,pagonabarraga2018,wagner, egleti2013}. In many of these cases the spontaneous polarization occurs in the absence of any explicit torques that act on the particles: No external torques occur when all external fields depend and act on position only, and no internal torques arise when the particles are  spheres.  
In equilibrium systems of spheres, the absence of torques implies local isotropy, and hence the emergence of nonzero local polarization is a genuine effect of nonequilibrium, as characterized by a nonzero spatially and orientationally resolved local one-body current $\textbf{J}.$

Important examples of these nonequilibrium situations include the spontaneous orientational ordering of ABPs against gravity in the sedimentation profile at large altitudes \cite{enculescu,ginot2018,mazza,schmidt2018}, the ordering upon adsorption against a (hard) wall \cite{speckjack, utrecht2018, wagner, egleti2013}, and the spontaneous polarization of the free interface between phase-separated active gas and liquid phases \cite{utrecht2018, schmidt2019, solon2018long,paliwal2016,paliwal2017}. There, $\textbf{M}$ points toward the active liquid in the case of purely repulsive particles \cite{utrecht2018, schmidt2019, solon2018long}, but toward the gas in the case of active Lennard-Jones particles \cite{paliwal2016,paliwal2017}. 
A range of different mechanisms and descriptions for the occurrence of the bulk phase separation has been put forward, such as, e.g.,\ kinetic blocking as a feedback mechanism \cite{cates2008,speck2016}, the existence of a nonequilibrium chemical potential \cite{utrecht2018,schmidt2019}, and effective interparticle attraction \cite{farage2015}.

The status of the nonequilibrium interface, however, has been claimed to be very different from what is known in equilibrium. Tailleur and coworkers \cite{solon2018short,solon2018long} find in their approach interface-to-bulk coupling, i.e.,\ the properties of the free interface affect the gas and liquid bulk states, which are in stable nonequilibrium coexistence.
Further, one can argue that due to the swim force, any nonvanishing polarization is necessarily associated with a one-body force distribution $\gamma s \textbf{M}$, where $\gamma$ is the translational friction constant and $s$ is the speed of free swimming. It is not inconceivable (and consistent with simple interface versus bulk dimensional analysis) that this force density compresses the phase toward which $\textbf{M}$ points at the expense of the other phase, and hence that it changes the properties of the coexisting phases.

\begin{figure}
\includegraphics[width=0.48 \textwidth]{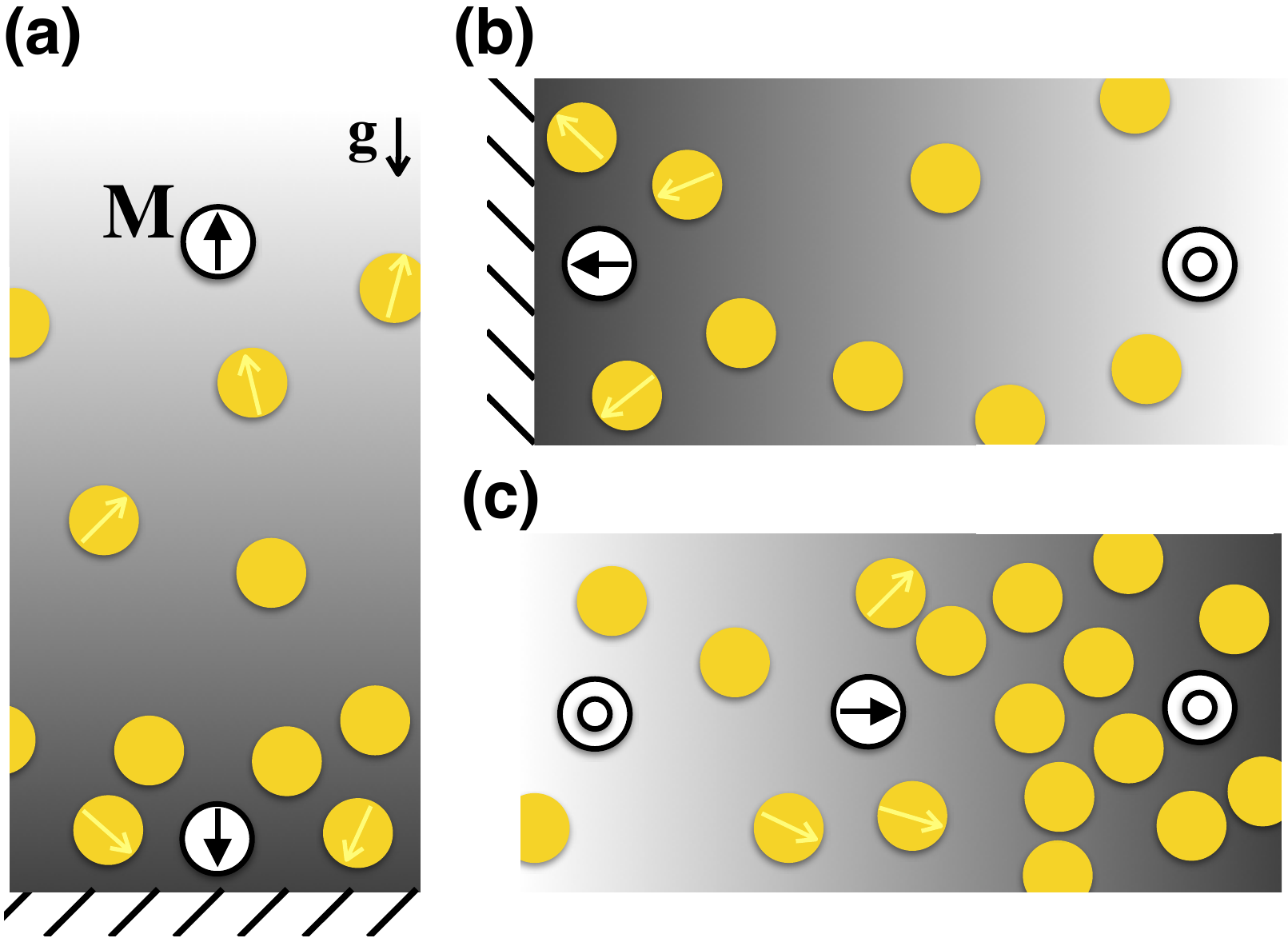}
\caption{\label{fig:1} Schematic of  systems in which the sum rules \eqref{eq:P0}, \eqref{eq:wall}, and \eqref{eq:mips} apply. (a) Sedimentation of ABPs under gravity $\textbf{g}$, which acts toward the lower confining wall (hatched area). (b) Adsorption of active particles at a semi-infinite wall (hatched area). (c) Motility induced phase separation of purely repulsive interacting ABPs. The yellow circles indicate active Brownian particles (ABPs) and the corresponding yellow arrows show exemplary particle orientations. The local particle polarization $\textbf{M}$ (white disks) can either vanish, indicated by a small circle, or point in a specific direction, indicated by a black arrow. The color gradient displays the density modulation from high values (dark) to low values (bright).} 
\end{figure}

Here we prove rigorously from first principles that the total interfacial polarization is a straightforward quantitative consequence of differing bulk currents in the coexisting phases and the rotational diffusion current $D_\text{rot}$. This rules out the total polarization as an underlying physical mechanism for the interface-to-bulk coupling \cite{solon2018long, solon2018short}.
Similarly, the total polarization of particles adsorbed at a wall is solely determined by $D_\text{rot}$ and the current in the corresponding bulk fluid, and thus constitutes a state function.
Furthermore, we show that in a system without explicit torques and with no total flux through the boundaries, the global orientational distribution function follows a free diffusion equation, so the global polarization vanishes in steady state; we also address the time-dependent case below. Figure \ref{fig:1} illustrates the three types of orientational ordering phenomena that we address in the following. Our derivation of the corresponding sum rules is based on the exact rotational equation of motion and on the continuity equation.

We describe ABPs on the level of their position- and orientation-resolved microscopic one-body density distribution $\rho(\textbf{r},\boldsymbol{\omega},t)$, where $\textbf{r}$ indicates position, $\boldsymbol{\omega}$ (unit vector) orientation, and $t$ time. Then the local polarization $\textbf{M}(\textbf{r},t)$ is a vector field defined as the first orientational moment of the density profile,
\begin{align}
 \textbf{M}(\textbf{r},t) &= \int \mathrm{d} \boldsymbol{\omega} \; \boldsymbol{\omega} \rho(\textbf{r},\boldsymbol{\omega},t), \label{eq:Ploc}
\end{align}
where the integral is over all orientations $\boldsymbol{\omega}$. 
The translational one-body current $\textbf{J}(\textbf{r},\boldsymbol{\omega},t)$ is the (microscopically resolved) measure of the direction and magnitude of the local flow of particles. 
As there are no explicit torques, the rotational motion is purely diffusive. Thus, the (in general) inhomogeneous density distribution $\rho$ generates a nonzero rotational current 
\begin{align}
\textbf{J}^\omega(\textbf{r},\boldsymbol{\omega},t)=-D_{\rm rot}\nabla^\omega \rho(\textbf{r},\boldsymbol{\omega},t),
\label{EQfreeRotationalDiffusion}
\end{align}
where $D_{\rm rot}$ is the rotational diffusion constant and $\nabla^\omega$ indicates the derivative with respect to orientation $\boldsymbol{\omega}$. As the dynamics evolve the microstates continuously in time and the total particle number $N$ remains constant, the one-body distributions satisfy the continuity equation,
\begin{align}
\dot\rho(\textbf{r},\boldsymbol{\omega},t) = -\nabla\cdot\textbf{J}(\textbf{r},\boldsymbol{\omega},t) - \nabla^\omega\cdot\textbf{J}^\omega(\textbf{r},\boldsymbol{\omega},t),
\label{EQcontinuity}
\end{align}
where $\dot{\rho}=\partial \rho/\partial t$ with $\dot{\rho}=0$ in steady state and $\nabla$ indicates the derivative with respect to position $\textbf{r}$. Note that the continuity equation \eqref{EQcontinuity} holds rigorously, independent of the presence and the type of interparticle interactions, particle-wall interactions, and external forces. The forces influence the translational and rotational motion, but not the form of \eqref{EQcontinuity}. All occurring terms in \eqref{EQcontinuity} can be sampled in computer simulations; see, e.g.,\ Ref.\ \cite{delasheras2019customflow}.

We first consider the total polarization for systems with vanishing total flux through the boundaries of volume $V$ at all times $t$, i.e.,\ $\int_{\partial V} \mathrm{d} \textbf{s} \cdot \textbf{J}(\textbf{r},\boldsymbol{\omega},t) = 0$, where $\mathrm{d} \textbf{s}$ denotes the vectorial surface element and $\partial V$ indicates the surface of volume $V$\!.
Here $V$ is arbitrary and can be chosen to be either the system volume, an enclosing larger volume that contains the system, or a subvolume of the system. The number of particles inside $V$ is $N = \int_V \mathrm{d} \textbf{r} \int \mathrm{d} \boldsymbol{\omega} \; \rho(\textbf{r},\boldsymbol{\omega},t)$.
We rewrite the spatially integrated density distribution as  $\int_V \mathrm{d} \textbf{r} \; \rho(\textbf{r}, \boldsymbol{\omega},t) = N f(\boldsymbol{\omega},t)$; this defines the global orientational distribution function $f(\boldsymbol{\omega},t)$, which is normalized at all times $t$, $ \int \mathrm{d} \boldsymbol{\omega} \; f(\boldsymbol{\omega},t) =1$.
Building the time derivative of the spatially integrated density distribution $\rho$ leads to
\begin{align}
N \dot{f}(\boldsymbol{\omega},t) &= \int\limits_V \mathrm{d} \textbf{r} \dot{\rho}(\textbf{r},\boldsymbol{\omega},t) \label{eq:a} \\
&= -\int\limits_V \mathrm{d} \textbf{r} \; \left( \nabla \cdot \textbf{J}(\textbf{r},\boldsymbol{\omega},t) + \nabla^{\omega} \cdot \textbf{J}^{\omega}(\textbf{r},\boldsymbol{\omega},t) \right), \label{eq:b}
\end{align}
where we used the continuity equation \eqref{EQcontinuity} to obtain \eqref{eq:b}. Assuming the absence of explicit torques and hence a free rotational diffusion current \eqref{EQfreeRotationalDiffusion}, applying the divergence theorem to the translational current contribution in \eqref{eq:b} yields 
\begin{align}
N \dot{f}(\boldsymbol{\omega},t)&= - \int\limits_{\partial V} \mathrm{d} \textbf{s} \cdot  \textbf{J}(\textbf{r},\boldsymbol{\omega},t) + \int\limits_V \mathrm{d} \textbf{r} \; D_\text{rot} \Delta^{\omega} \rho(\textbf{r},\boldsymbol{\omega},t) \label{eq:c} \\
&= D_\text{rot} N \Delta^{\omega}  f(\boldsymbol{\omega},t), \label{eq:d}
\end{align}
 where the orientational Laplace operator is $\Delta^{\omega} = \nabla^{\omega} \cdot \nabla^{\omega}$.
The first term on the right hand side of Eq.\ \eqref{eq:c} vanishes due to the vanishing flux boundary condition and the second term can be rewritten as \eqref{eq:d} using the definition of $f$.
Dividing Eq.\ \eqref{eq:d} by the particle number $N$ yields a free diffusion equation for the orientational distribution function
\begin{align}
\dot{f}(\boldsymbol{\omega},t) &= D_\text{rot} \Delta^{\omega}  f(\boldsymbol{\omega},t).\label{eq:freeDiff}
\end{align}
Note that Milster et al. \cite{sokolov2017} derived an equation similar to Eq.\ \eqref{eq:freeDiff} for the orientational distribution function of two-dimensional ABPs with negligible translational diffusion.
We consider the system to be in steady state, $\dot{\rho}(\textbf{r},\boldsymbol{\omega},t) = 0$, and thus also $\dot{f}(\boldsymbol{\omega},t) = 0$, which simplifies the diffusion equation \eqref{eq:freeDiff} to 
\begin{align}
\Delta^{\omega} f(\boldsymbol{\omega}) =0. \label{eq:stst}
\end{align}
The only solutions in two and three dimensions (2D and 3D) of Eq.\ \eqref{eq:stst} are constants, $f=(2\pi)^{-1}$ for two-dimensional systems and $f=(4\pi)^{-1}$ in three dimensions. Hence, we conclude \cite{footnote1} that the global orientational distribution function $f$ is independent of $\boldsymbol{\omega}$ and the total polarization $\textbf{M}_\text{tot}$ vanishes,
\begin{align}
\textbf{M}_\text{tot} &= \int\limits_V \mathrm{d} \textbf{r} \; \textbf{M}(\textbf{r}) = 0. \label{eq:P0}
\end{align}
In practice, the result \eqref{eq:P0} can be used as a consistency check in computer simulations and in theoretical descriptions.
It is trivially satisfied in equilibrium systems without explicit torques, as such systems imply local isotropy and hence local and total polarization both vanish.

We emphasize that Eq.\ \eqref{eq:P0} holds in all steady states, independent of the existence of external potentials or the present type of interparticle or particle-wall interactions, in each (sub)volume $V$ with zero net flux through its boundaries.
The local and hence also the total fluxes through the surface of the considered volume are zero if the orientational distribution function is homogeneous at the surface and the current can be expressed as $\textbf{J}(\textbf{r},\boldsymbol{\omega}) = J_b \boldsymbol{\omega}$. The magnitude $J_b$ is equal to the first Fourier coefficient of the current. The condition for the current is satisfied, e.g.,\ in isotropic bulk states or in regions of vanishing current.

As a first application of Eq.\ \eqref{eq:P0}, we consider sedimentation of ABPs \cite{enculescu,solon2015}. Figure \ref{fig:1}(a) illustrates two prominent effects that occur: 
(i) The particle orientation points toward the lower confining wall \cite{enculescu, mazza, schmidt2018, ginot2018}. This leads to a particle accumulation at the wall on top of the effect of gravity and may be interpreted as a self trapping mechanism. 
(ii) The sedimentation length increases as compared to passive particles due to the alignment of the swimmers against gravity $\textbf{g}$ at large distances from the wall \cite{enculescu, mazza, schmidt2018, ginot2018}. Both effects can be interpreted as originating from a dynamical balance between the spatial self-sorting of the active particles and the counteracting mechanism of rotational diffusion.
An upward oriented particle for example swims on average toward higher altitudes until its orientation changes via particle rotation. 

Each of the above phenomena (i) and (ii) generates nonvanishing local polarization \cite{enculescu,mazza, schmidt2018, ginot2018} [cf.\ Fig. \ref{fig:1}(a)] and both have at first sight no relationship with each other. But as the flux through the boundaries is zero \cite{footnote2}, the total polarization has to vanish in steady state, cf.\ \eqref{eq:P0}. 
Thus, if the volume $V$ is divided into bottom and top subvolumes, both partial polarizations have to cancel each other, independent of the division itself.
This effect is non-local as the accumulation and polarization at the bottom determines the overall particle orientation in the remaining volume. 

We next consider non-vanishing total flux through the boundaries. We derive a spatially resolved (``local'') version of the sum rule for the ubiquitous two-dimensional system in steady state. 
In two dimensions, the orientation vector can be written as $\boldsymbol{\omega} = (\cos \varphi, \sin \varphi)$, where $\varphi$ is the angle measured against the positive $x$-axis, and the orientational derivative $\nabla^\omega$ reduces to $\partial/\partial \varphi$. 
We assume, as a relevant case, translational invariance along the $y$-axis. (Note, however, that this restriction is not necessary \cite{footnote3}).
Because of the assumption of translational invariance, the density $\rho(x,\varphi)$ and $x$-component of the current $J^x(x,\varphi)$ are even in the angle $\varphi$ as both are invariant under reflection at $x$-axis, $y \to -y$ and $\varphi \to - \varphi$ \cite{schmidt2019}. 
So the angular Fourier decomposition of both quantities consists only of cosines. The density thus may be expressed as
\begin{align}
\rho(x,\varphi) = \sum\limits_{n=0}^{\infty} \rho_{n}(x) \cos(n \varphi), \label{eq:rhoFourier}
\end{align}
where $\rho_n(x)$ indicates the $n$th Fourier coefficient of the density profile.
Using \eqref{eq:rhoFourier} in the expression for the polarization \eqref{eq:Ploc} yields $\textbf{M}=(\pi \rho_1(x),0)$. The $y$-component $M_y$ vanishes due to the symmetry of the density distribution \eqref{eq:rhoFourier}, so the magnitude of the polarization is equal its $x$-component, $M = M_x$. 
The $x$-component of the current can be Fourier decomposed similarly as
\begin{align}
J^x(x,\varphi) = \sum\limits_{n=0}^{\infty} J^x_n(x) \cos(n \varphi), \label{eq:Fourier}
\end{align}
where $J^x_n(x)$ denotes the $n$th Fourier coefficient and thus the $n$th orientational moment of the current. 
As the rotational current consists only of the thermal free diffusion contribution \eqref{EQfreeRotationalDiffusion}, the continuity equation \eqref{EQcontinuity} simplifies for steady states to 
\begin{align}
\frac{\partial J^x(x,\varphi)}{\partial x} = D_\text{rot} \frac{\partial^2 \rho(x,\varphi)}{\partial \varphi^2}. \label{eq:continuity}
\end{align}
Equation \eqref{eq:continuity} is satisfied, e.g.,\ for the case of motility induced phase separation \cite{schmidt2019}. 
Insertion of the Fourier decomposition \eqref{eq:Fourier} in Eq.\ \eqref{eq:continuity} and integrating twice in the angle $\varphi$ allows us to solve the equation for density $\rho$. Evaluation of both indefinite integrals leads to
\begin{align}
\rho(x,\varphi) &= -\frac{1}{D_\text{rot}} \sum\limits_{n=1}^{\infty} \frac{\partial J_n^x(x)}{\partial x} \frac{\cos (n \varphi)}{n^2} + \rho_0, \label{eq:rho}
\end{align}
where we have used the Fourier expansion of the current \eqref{eq:Fourier} and the integration constant $\rho_0$ indicates the average density, i.e.,\ the total number of particles per system volume and per radians.
The integration constant of the first integral vanishes, since a linear $\varphi$-term does not satisfy the $2\pi$-periodicity in angle $\varphi$.

The polarization profile $M(x)$ \eqref{eq:Ploc} can be simplified as $M(x)= \int_{0}^{2\pi} \mathrm{d} \varphi \; \rho(x,\varphi) \cos \varphi$ using the present symmetries. Inserting the expansion of $\rho$ \eqref{eq:rho} and evaluating the integral over all orientations yields
\begin{align}
M(x)= -\frac{\pi}{D_\text{rot}} \frac{\partial J_1^x(x)}{\partial x}. \label{eq:Px}
\end{align}
That is for each position $x$ the local polarization is proportional to the spatial change in the first moment of the current. The spatially resolved relation \eqref{eq:Px} constitutes a local sum rule, similarly determined by Refs. \cite{utrecht2018,schmidt2019} in the special case of ABPs. The derivation here is more general and based only on the continuity equation with freely diffusive rotational motion. 

In order to derive a global sum rule, we spatially integrate the  exact local sum rule \eqref{eq:Px},
\begin{align}
M_\text{tot} = \int\limits_{x_1}^{x_2}  \mathrm{d} x \int\limits_{y_1}^{y_2} \mathrm{d} y \; M(x)  = \frac{\pi L_y}{D_\text{rot}} \left( J_1^x(x_1) - J_1^x(x_2) \right), \label{eq:e1}
\end{align} 
which determines the total polarization $M_\text{tot}$ in the integration volume $V$. For simplicity we restrict ourself to rectangular areas $V$ aligned with the coordinate axes. The integration limits are set to the arbitrary positions $x_1$ and $x_2$ for the $x$-coordinate and  $y_1$ and $y_2$ for the $y$-coordinate. Because of the translational invariance, the $y$-integral can  be explicitly evaluated and gives the length of $y$-integration, $L_y = y_2 - y_1$. 
In the following we thus consider the total polarization per unit length in the $y$-direction, $M_\text{tot}/L_y$.

Equation \eqref{eq:e1} holds for ABPs in a large variety of situations. We address two general relevant cases in the following. First we consider ABPs absorbed at a (hard or soft) planar wall [see figure \ref{fig:1}(b)]. We set a wall parallel to the $y$-axis at $x=0$.  As the density vanishes inside the wall, the one-body current $\textbf{J}(x) = 0$ for $x \rightarrow -\infty$. 
For $x\rightarrow\infty$, the semi-infinite system approaches an isotropic bulk fluid, so the current is $\textbf{J}(x) = J_b \boldsymbol{\omega}$, due to symmetry. 
The (constant) magnitude of the bulk current, $J_{b}$, equals the first Fourier component, $J_b = J_1^x$. 
Setting the limits of integration in Eq.\ \eqref{eq:e1} to $x_1 \rightarrow -\infty$ and $x_2 \rightarrow \infty$ and using the known expressions for the currents, simplifies the total polarization at the wall per unit $y$-length to
\begin{align}
\frac{M_\text{tot}}{L_y} = - \frac{\pi}{D_\text{rot}} J_b. \label{eq:wall}
\end{align} 
Hence, the absolute value of $\textbf{M}_\text{tot}$ is solely determined by the bulk current and the rotational diffusion constant. Recall that $\textbf{M}_\text{tot}$ is oriented along the $x$-axis due to the translational symmetry, so the sign of the right-hand side of Eq.\ \eqref{eq:wall} determines whether the total polarization points toward or against the wall. As the free swim speed $s\geq0$, $J_{b}$ is greater or equal to zero, and hence the total polarization points towards the wall. (Note that interparticle interactions only tend to reduce the absolute value of the bulk current due to drag effects). 
Because of the global sum rule \eqref{eq:wall}, the sign of the total $x$-polarization per unit length is  negative. A vanishing bulk current $J_{b}=0$ constitutes a special case, which leads to a vanishing total polarization as one would expect to occur for a system of passive spheres.
We conclude that the total swim force density $\int \mathrm{d} \textbf{r} \mathrm{d} \boldsymbol{\omega} s \gamma \rho \boldsymbol{\omega}$ always points toward the wall, so that the total polarization also points to the wall (except if the total polarization vanishes). The direction of the total polarization is hence independent of both the particle-wall and the interparticle interaction.
Furthermore, due to locality of both interactions the bulk itself, in particular the bulk current $J_b$, is independent of the wall. Thus, the total polarization $\textbf{M}_\text{tot}$ only depends on bulk quantities via Eq.\ \eqref{eq:wall} and constitutes a state function. This extends the work of Tailleur and coworkers \cite{solon2015A, solon2015B}, who investigated whether pressure is a state function in active fluids. Note that the magnitude and structure of the local polarization \textit{profile} $M(x)$ may depend on both the wall-particle and the interparticle interaction potentials. 

As a second relevant example, we consider the phase separation of ABPs [schematic sketch in Fig. \ref{fig:1}(c)]. The particles phase separate in a dense (liquid) and a dilute (gas) bulk fluid. Since both coexisting bulk states are isotropic, the coexisting bulk currents are proportional to the orientation $\boldsymbol{\omega}$ and the corresponding magnitudes are $J_g$ in the gas and $J_l$ in the liquid bulk phase.
Using those relations for the bulk current and setting the limits of integration inside an isotropic bulk phase, i.e.,\ $x_1\rightarrow -\infty$ and $x_2 \rightarrow \infty$, simplifies Eq.\ \eqref{eq:e1} to
\begin{align}
\frac{M_\text{tot}}{L_y} = \frac{\pi}{ D_\text{rot}} \left( J_{g} - J_{l} \right). \label{eq:mips}
\end{align}
Hence the difference between both local bulk currents, scaled with the rotational diffusion constant,  determines the total polarization per transversal length. Equation \eqref{eq:mips} constitutes an exact global sum rule.
For particles interacting via the Weeks-Chandler-Anderson potential, which is a Lennard-Jones potential cut and shifted at its minimum to be purely repulsive, the swimmers align towards the denser phase in the interfacial region \cite{utrecht2018, schmidt2019}; cf.\ Fig. \ref{fig:1}(c). Hence, the total polarization is also directed towards the dense phase and it is positive.  According to Eq.\ \eqref{eq:mips}, one expects a higher current in the dilute phase in comparison to the dense phase, which is in qualitative and quantitative agreement with simulation data \cite{schmidt2019,footnote4}. 
In contrast, for active Lennard-Jones particles, the total polarization was found to point toward the dilute phase \cite{paliwal2017}.
A sketch of the system would be similar to Fig. \ref{fig:1}(c), but with an inverted polarization arrow. Using the total polarization to calculate the difference between both bulk currents from the global sum rule \eqref{eq:mips}, we predict a higher current in the liquid than in the gas.
Note that the particle polarization is primarily located at the interface, since the polarization in bulk vanishes due to isotropy; cf.\ Eq.\ \eqref{eq:e1}. 

A physical interpretation of the global sum rule \eqref{eq:mips} is that the interfacial quantity $M_\text{tot}$ is solely determined by the bulk values $J_g$ and $J_l$. In other words, the interfacial polarization is a mere consequence of the properties of the bulk states. This interpretation follows from the locality of the short-ranged interparticle interactions, which is a similar reasoning as in the case of particles in front of a semi-infinite wall [cf. Fig \ref{fig:1}(b)].
The combination of the expression \eqref{eq:mips} and the locality of interparticle interactions lets the non-local influence of $\textbf{M}_\text{tot}$ on the entire bulk seem implausible.  
This questions the conclusion of Solon et al. (p. 16, \cite{solon2018long}) that ``the phase coexistence densities [...] is controlled by the polar ordering of particles at the gas–liquid interface.'' 
It seems more reasonable that the interface is a consequence of the bulk and not vice versa, especially since no mechanism has been identified which would generate these non-local effects. Furthermore, our interpretation is in agreement with the theory of Ref. \cite{schmidt2019} where no interfacial contributions are required to describe the bulk and the gas-liquid coexistence, as is the case in equilibrium.

We next generalize the steady state relationship \eqref{eq:P0} and
consider the time dependence of $\textbf{M}_\text{tot}$. We restrict
ourselves to cases of vanishing total flux through the boundaries of
the considered volume $V$ at all times. The time-dependent total
polarization $\textbf{M}_\text{tot}(t)$ is then given as
\begin{align}
\textbf{M}_\text{tot}(t) = \int\limits_V \mathrm{d} \textbf{r} \; \textbf{M}(\textbf{r},t) = N \int \mathrm{d} \boldsymbol{\omega} \;  \boldsymbol{\omega} f(\boldsymbol{\omega},t), \label{eq:Mtot}
\end{align} 
and thus can be determined via the global orientational distribution
function $f(\boldsymbol{\omega},t)$. We first consider two-dimensional
systems. Hence, as above, the orientation vector is
$\boldsymbol{\omega} = (\cos\varphi,\sin\varphi)$, where $\varphi$ is
an angular coordinate and $\Delta^\omega$ simplifies to $\partial^2 /
\partial \varphi^2$. So, $f(\boldsymbol{\omega},t)$ is given as the
solution of Eq.\ \eqref{eq:freeDiff},
\begin{align}
f(\varphi,t) = \sum_{n=0}^\infty \left( a_n \cos(n \varphi) +  b_n \sin(n \varphi) \right) e^{-n^2 D_\text{rot} t}, \label{eq:f2D}
\end{align}
where the constants $a_n$, $b_n$ are determined by the initial
conditions. Inserting the global orientational distribution function
\eqref{eq:f2D} into Eq.\ \eqref{eq:Mtot} and carrying out the angular
integral yields the temporal behavior of the total polarization as an
exponential decay,
\begin{align}
\textbf{M}_\text{tot}(t) = \left(\begin{array}{c}a_1\\b_1\end{array}\right) e^{-D_\text{rot}t}, \label{eq:Mtot2D}
\end{align}
where $1/D_\text{rot}$ is the time constant and the vector $(a_1,b_1)$
is the initial polarization at time $t=0$.

In three spatial dimensions, we parametrize $\boldsymbol{\omega} =
(\sin\theta\cos\varphi,\sin \theta\sin\varphi,\cos\theta)$, where
$\theta$ and $\varphi$ indicate polar and azimuthal angles. Then
Eq.\ \eqref{eq:freeDiff} is solved by
\begin{align}
f(\theta, \varphi,t) = \sum\limits_{l=0}^\infty \sum\limits_{m=-l}^l a_{lm} Y_{lm}(\theta, \varphi) e^{-\frac{D_\text{rot}}{l(l+1)} t}, \label{eq:f3D}
\end{align}
where the constants $a_{lm}$ are again set by initial conditions and $Y_{lm}(\theta,\varphi)$ indicate the spherical harmonics. Insertion of \eqref{eq:f3D} into Eq.\ \eqref{eq:Mtot} gives 
\begin{align}
\textbf{M}_\text{tot}(t) =  \sum\limits_{l=0}^\infty \textbf{M}_l e^{-\frac{D_\text{rot}}{l(l+1)} t}, \label{eq:Mtot3D}
\end{align}
where we have defined the constants $\textbf{M}_l = \int \mathrm{d}
\boldsymbol{\omega} \sum_{m} a_{lm} Y_{lm}(\theta,\varphi)
\boldsymbol{\omega}$.  Hence, in both the 2D and 3D cases,
$\textbf{M}_\text{tot}$ decays exponentially in time. The dynamics
depend only on the rotational diffusion constant and on the initial
conditions. Clearly in the limit $t\to\infty$, the results for the time
dependence of $\textbf{M}_\text{tot}$, \eqref{eq:Mtot2D} and
\eqref{eq:Mtot3D}, reduce to the steady state sum rule \eqref{eq:P0}
of vanishing total polarization.

Furthermore, one can extend the obtained sum rules to higher (e.g.,\ nematic) order moments, e.g.,\ $M_n(\textbf{r}) = \int_0^{2\pi} \mathrm{d} \varphi \cos(n\varphi) \rho(\textbf{r},\varphi)$ in two dimensional systems. For a vanishing flux through the surface of the volume, those higher moments in the considered volume are all equal to zero in steady state, as is the polarization [cf.\ Eq.\ \eqref{eq:P0}], and their time evolution can be derived from Eqs. \eqref{eq:f2D}, \eqref{eq:f3D}.
In translationally invariant two-dimensional systems the sum rules are similar to Eqs. \eqref{eq:Px} and \eqref{eq:e1}, where the $n$th moment $M_n$ corresponds to the spatial derivative of the $n$th moment of the current $J_n^x$. Since higher moments of the bulk current $J_{n>1}^x$ vanish in bulk due to symmetry, the total higher order moments $M^\text{tot}_{n>1}$ are also zero for particle adsorption at a wall or motility induced phase separation. Hence, these total moments cannot contribute to determine the bulk densities. Note, however, that the local structure of these higher order moments is non-trivial in general.

To conclude, we have demonstrated that polarization and current distribution of ABPs are intimately connected. Using the continuity equation, together with the properties of free rotational diffusion, we have derived three exact global sum rules \eqref{eq:P0}, \eqref{eq:wall}, and \eqref{eq:mips}. These imply, respectively, (i) that the total system polarization vanishes, (ii) that the polarization at a wall is determined by the bulk current and hence represents a state function, and (iii) that for phase-separated fluid states the polarization of the free interface is given by the difference of bulk current in the coexisting active bulk phases. Note that Eq.\ \eqref{eq:mips} is indeed satisfied qualitatively and quantitatively in the theory of Refs. \cite{schmidt2019,tension2019}. These global sum rules, as well as the local sum rule \eqref{eq:Px}, can be useful as consistency checks for simulations and theories and can also be used as an input for theoretical descriptions. 
One could apply the derived local and global sum rules to further interesting systems: The relations hold in case of spatial inhomogeneous activity $s(\textbf{r})$ as considered by Sharma et al. \cite{brader2017} and Hasnain et al. \cite{menzl2017} or for spatially varying translational diffusion \cite{menzl2017}, as long as the rotational diffusion coefficient is kept constant.

We thank D. de las Heras for stimulating discussions and critical reading of the manuscript.


\begin{thebibliography}{31}

\bibitem{marchetti2013} M. C. Marchetti, J. F. Joanny, S. Ramaswamy, T. B. Liverpool, J. Prost, M. Rao, and R. A. Simha, Rev. Mod. Phys. \textbf{85}, 1143 (2013).

\bibitem{review2} P. Romanczuk, M. B{\"a}r, W. Ebeling, B. Lindner, and L. Schimansky-Geier, Eur. Phys. J. Spec. Top. \textbf{202}, 1 (2012).

\bibitem{review3} C. Bechinger, R. Di Leonardo, H. L{\"o}wen, C. Reichhardt, G. Volpe, and G. Volpe, Rev. Mod. Phys. \textbf{88}, 045006 (2016).

\bibitem{schmidt2018} S. Hermann and M. Schmidt, Soft Matter \textbf{14}, 1614 (2018).

\bibitem{enculescu} M. Enculescu and H. Stark, Phys. Rev. Lett. \textbf{107}, 058301 (2011).

\bibitem{mazza} J. Vachier and M. G. Mazza, Eur. Phys. J. E \textbf{42}, 11 (2019).

\bibitem{ginot2018} F. Ginot, A. Solon, Y. Kafri, C. Ybert, J. Tailleur, and C. Cottin-Bizonne, New J. Phys. \textbf{20}, 115001 (2018).

\bibitem{stark2014} M. Hennes, K. Wolff, and H. Stark, Phys. Rev. Lett. \textbf{112}, 238104 (2014).

\bibitem{pagonabarraga2018} A. Mart{\'i}n-G{\'o}mez, D. Levis, A. D{\'i}az-Guilera and I. Pagonabarraga, Soft Matter \textbf{14}, 2610 (2018).

\bibitem{speckjack} T. Speck and R. L. Jack, Phys. Rev. E \textbf{93}, 062605 (2016).

\bibitem{wagner} C. G. Wagner, M. F. Hagan, and A. Baskaran, J. Stat. Mech. \textbf{2017}, 043203 (2017).

\bibitem{egleti2013} J. Elgeti and G. Gompper, Europhys. Lett. \textbf{101}, 48003 (2013).

\bibitem{utrecht2018} S. Paliwal, J. Rodenburg, R. van Roij, and M. Dijkstra, New J. Phys. \textbf{20}, 015003 (2018).

\bibitem{schmidt2019} S. Hermann, P. Krinninger, D. de las Heras, and M. Schmidt, Phys. Rev. E \textbf{100}, 052604 (2019).

\bibitem{solon2018long} A. P. Solon, J. Stenhammar, M. E. Cates, Y. Kafri, and J. Tailleur, New J. Phys. \textbf{20}, 075001 (2018).

\bibitem{paliwal2016} V. Prymidis, S. Paliwal, M. Dijkstra, and L. Filion, J. Chem. Phys. \textbf{145}, 124904 (2016).

\bibitem{paliwal2017} S. Paliwal, V. Prymidis, L. Filion, and M. Dijkstra, J. Chem. Phys. \textbf{147}, 084902 (2017).

\bibitem{cates2008} J. Tailleur and M. E. Cates, Phys. Rev. Lett. \textbf{100}, 218103 (2008).

\bibitem{speck2016} D. Richard, H. L{\"o}wen, and T. Speck, Soft Matter \textbf{12}, 5257 (2016).

\bibitem{farage2015}
  T.~F.~F. Farage, P. Krinninger, and J.~M. Brader, Phys. Rev. E {\bf 91}, 042310 (2015).

\bibitem{solon2018short} A. P. Solon, J. Stenhammar, M. E. Cates, Y. Kafri, J. Tailleur, Phys. Rev. E \textbf{97}, 020602(R) (2018).

\bibitem{delasheras2019customflow} D. de las Heras, J. Renner, and M. Schmidt, Phys. Rev. E {\bf 99}, 023306 (2019). 

\bibitem{sokolov2017} See Appendix A of S. Milster, J. N{\"o}tel, I. M. Sokolov, and L. Schimansky-Geier, Eur. Phys. J. Spec. Top. \textbf{226}, 2039 (2017).

\bibitem{footnote1}
In order to show that Eq. \eqref{eq:P0} holds, we calculate the total polarization, which is defined as the spatial integral of the local polarization \eqref{eq:Ploc}, $\textbf{M}_\text{tot} = \int_V \mathrm{d} \textbf{r} \; \textbf{M}(\textbf{r}) = \int \mathrm{d} \boldsymbol{\omega} \; \boldsymbol{\omega} \int_V \mathrm{d} \textbf{r} \; \rho(\textbf{r},\boldsymbol{\omega})$, where the spatial and orientational integral were interchanged. The spatially integrated density can be expressed as $N f$, which is constant for two and three dimensional systems in case of steady state and a vanishing total flux through the boundaries, cf.\ Eq. \eqref{eq:stst}. Thus, the total polarization simplifies to  $\textbf{M}_\text{tot} = N f \int \mathrm{d} \boldsymbol{\omega} \; \boldsymbol{\omega} = 0$.

\bibitem{solon2015} A. P. Solon and M. E. Cates, Eur. Phys. J. Spec. Top. \textbf{224}, 1231 (2015).

\bibitem{footnote2} 
The density and hence the current through the top and the bottom of the system vanish due to gravity and in the lower confining wall, respectively. The flux through the left system sides cancels with those through the right due to translational invariance. Hence, the total flux though the surface of the system is zero.

\bibitem{footnote3}
If one discards the translational symmetry along the $y$-axis, odd and even functions contribute to the density and the current. The Fourier decompositions of the density is then $\rho(x,y,\varphi) = \sum_{n=0}^{\infty} \rho^c_{n}(x,y) \cos(n \varphi) + \rho^s_{n}(x,y) \sin(n \varphi)$, where $\rho^c_{n}$ indicates the $n$th Fourier coefficient of the cosine contributions and $\rho^s_{n}$ denote the coefficients of the sine contributions. Similarly the Fourier expansion of the current is $\textbf{J}(x,y,\varphi) = \sum_{n=0}^{\infty} \textbf{J}^c_n(x,y) \cos(n \varphi)+\textbf{J}^s_n(x,y) \sin(n \varphi)$, where $\textbf{J}^c_n$ corresponds to the $n$th cosine and $\textbf{J}^s_n$ to the $n$th sine Fourier coefficient of the current.
The generalized expression of the local polarization profile \eqref{eq:Px} is then $M_x =\pi \rho_1^c =\pi \left( \partial J^{x,c}_1/\partial x + \partial J^{y,c}_1/\partial y \right)/ D_\text{rot}$ for the $x$-component and $M_y =\pi \rho_1^s =\pi \left( \partial J^{x,s}_1/\partial x + \partial J^{y,s}_1/\partial y \right)/ D_\text{rot}$ for the $y$-component.

\bibitem{solon2015A} A. P. Solon, J. Stenhammar, R. Wittkowski, M. Kardar, Y. Kafri, M. E. Cates, and J. Tailleur, Phys. Rev. Lett. \textbf{114}, 198301 (2015).

\bibitem{solon2015B} A. P. Solon, Y. Fily, A. Baskaran, M. E. Cates, Y. Kafri, M. Kardar, and J. Tailleur, Nat. Phys. \textbf{11}, 673 (2015).

\bibitem{footnote4} As an example, based on the simulation data
  of Ref. \cite{schmidt2019} for motility-induced phase separation, we
  obtain the total polarization, i.e.,\ the left-hand side of
  Eq.\ \eqref{eq:mips}, as $3.77/\sigma$. This agrees well with the
  value $3.73/\sigma$, which we obtain for the right-hand side of
  Eq.\ \eqref{eq:mips}. Here $\sigma$ indicates the particle size; for
  further simulation details, see Ref.\ \cite{schmidt2019}. 

\bibitem{tension2019} S. Hermann, D. de las Heras, and M. Schmidt, Phys. Rev. Lett. \textbf{128}, 268002 (2019).

\bibitem{brader2017} A. Sharma and J. M. Brader, Phys. Rev. E \textbf{96}, 032604 (2017).

\bibitem{menzl2017} J. Hasnain, G. Menzl, S. Jungblut, and C. Dellago, Soft Matter \textbf{13}, 930 (2017).

\end{thebibliography}
\end{document}